\documentclass[aps,prd,reprint,superscriptaddress,nofootinbib]{revtex4-1}
\usepackage{graphicx}
\usepackage{dcolumn}
\usepackage{bm}
\usepackage{epstopdf}
\usepackage{amsmath}
\usepackage{amssymb}
\usepackage{hyperref}
\usepackage{color}
\usepackage{cancel}
\usepackage{braket}
\usepackage{url}

\begin{document}

\raisebox{20pt}[0pt][0pt]{\hspace*{0mm} RIKEN-QHP-514, RIKEN-iTHEMS-Report-22, YITP-22-01}

\title{Optimized Two-Baryon Operators in Lattice QCD}

\author{Yan Lyu
}
\email{helvetia@pku.edu.cn}
\affiliation{State Key Laboratory of Nuclear Physics and Technology, School of Physics, Peking University, Beijing 100871, China }
\affiliation{Quantum Hadron Physics Laboratory, RIKEN Nishina Center, Wako 351-0198, Japan}
\author{Hui Tong
}
\affiliation{State Key Laboratory of Nuclear Physics and Technology, School of Physics, Peking University, Beijing 100871, China }
\affiliation{Interdisciplinary Theoretical and Mathematical Sciences Program (iTHEMS), RIKEN, Wako 351-0198, Japan}
\author{Takuya Sugiura
}
\affiliation{Interdisciplinary Theoretical and Mathematical Sciences Program (iTHEMS), RIKEN, Wako 351-0198, Japan}

\author{Sinya Aoki
}
\email{saoki@yukawa.kyoto-u.ac.jp}
\affiliation{Center for Gravitational Physics, Yukawa Institute for Theoretical Physics,
 Kyoto University, Kyoto 606-8502, Japan}
 \affiliation{Quantum Hadron Physics Laboratory, RIKEN Nishina Center, Wako 351-0198, Japan}
\author{\\ Takumi Doi
}
\email{doi@ribf.riken.jp}
\affiliation{Quantum Hadron Physics Laboratory, RIKEN Nishina Center, Wako 351-0198, Japan}
\affiliation{Interdisciplinary Theoretical and Mathematical Sciences Program (iTHEMS), RIKEN, Wako 351-0198, Japan}
\author{Tetsuo Hatsuda
}
\email{thatsuda@riken.jp}
\affiliation{Interdisciplinary Theoretical and Mathematical Sciences Program (iTHEMS), RIKEN, Wako 351-0198, Japan}
\author{Jie Meng
}
\email{mengj@pku.edu.cn}
\affiliation{State Key Laboratory of Nuclear Physics and Technology, School of Physics, Peking University, Beijing 100871, China }
\affiliation{Yukawa Institute for Theoretical Physics, Kyoto University, Kyoto 606-8502, Japan}
\author{Takaya Miyamoto
}
\affiliation{Quantum Hadron Physics Laboratory, RIKEN Nishina Center, Wako 351-0198, Japan}


\begin{abstract}
A set of optimized interpolating operators which are dominantly coupled to each eigenstate of two baryons on the lattice is constructed by the HAL QCD method. 
To test its validity, we consider heavy dibaryons $\Omega_{3Q}\Omega_{3Q}$ ($Q=s,c$) calculated by (2+1)-flavor lattice QCD simulations with nearly physical pion mass.
The optimized two-baryon operators are shown to provide effective energies of the ground and excited states separately stable as a function of the Euclidean time.
Also they agree to the eigenenergies in a finite lattice box obtained from the leading-order HAL QCD potential  $V({\bm r})$ within statistical errors.
The overlapping factors between the optimized sink operators and the state created by the wall-type source operator indicate that $V({\bm r})$ can be reliably extracted, no matter whether the spacetime correlation of two baryons is dominated by the ground state or the excited state. 
It is suggested that the optimized set of operators is useful for variational studies of hadron-hadron interactions.
\end{abstract}

\maketitle



\section{Introduction}
Accurate determination of hadron-hadron interactions from QCD is one of the most challenging problems in nuclear and particle physics~\cite{Drischler2019,Aoki2020}.  
To achieve the goal, two theoretical methods have been proposed in lattice QCD (LQCD); the L\"uscher's finite volume method~\cite{Luscher1991} and the HAL QCD method~\cite{Ishii2007,Aoki2010,Ishii2012}.  
The former focuses on the temporal behavior of the hadronic correlations, from which the scattering phase shifts are extracted through the  L\"uscher's finite volume formula.
On the other hand,  the latter considers the spacetime behavior of the hadronic correlations, from which physical observables are extracted through the equal-time Nambu-Bethe-Salpeter (NBS) amplitude, or the NBS wave function in short. 
   
Although these two methods are like looking at the two sides of the same coin and are theoretically related with each other, it has been realized that the  ``naive"  plateau fitting in the L\"uscher's finite volume method without variational analysis sometimes leads to misleading results for two-baryon systems.
It originates from the fact that separating the ground state from excited states of two baryons below inelastic threshold is exponentially difficult on the lattice as the splitting in energy becomes zero in the large volume
limit~\cite{Lepage1989}: Detailed account of this issue is given in Refs.~\cite{Iritani2016,Iritani2019Jhep} (see also, Refs.~\cite{Francis2019,Horz2021,Amarasinghe2021}). 

In contrast, the time-dependent version of the HAL QCD method~\cite{Ishii2012} does not need to identify each level, since the same NBS integral kernel (i.e. the energy-independent non-local potential) governs all the elastic scattering states simultaneously.
In other words, any superposition of the NBS wave functions of different energies below inelastic threshold leads to the same non-local potential.
By using this nice property, one can construct a set of optimized interpolating operators and make a firm connection between the L\"uscher's finite volume method and the HAL QCD method for two baryons, as demonstrated in the (2+1)-flavor LQCD simulations with heavy pion mass, $m_{\pi} \simeq$ 510 MeV,  and the lattice volumes, (3.6, 4.3, 5.8 fm)$^3$~\cite{Iritani2019Jhep}.
Such optimized operators can also be used to check the validity of the derivative expansion of the non-local potential in the HAL QCD method. 
   
The purpose of this paper is to further explore the idea of optimized operators proposed in Ref.~\cite{Iritani2019Jhep} and study whether it is applicable to (2+1)-flavor LQCD near the physical pion mass, $m_{\pi} \simeq$ 146 MeV, with a large lattice volume, (8.1 fm)$^3$. 
We consider two heavy dibaryons $\Omega_{3Q}\Omega_{3Q}$ in the $^1S_0$ channel with $Q=s$ and $c$, where $\Omega_{3Q}$ implies the spin-3/2 baryon composed of three valence quarks with flavor $Q$. 
Both systems are in the unitary regime with large scattering lengths as demonstrated recently by the $(2+1)$-flavor LQCD simulations~\cite{Gongyo2018,Lyu2021}.
The statistical errors in these systems are relatively small in comparison to the baryons with light valence quarks, so that one can make quantitative analysis on the effect of the optimized operators as well as on the uncertainty due to derivative expansion of the non-local potential.

This paper is organized as follows.
After a brief review of the HAL QCD method  in Sec.~\ref{Sec_II}, we introduce a general framework to define the optimized two-baryon operators through the HAL QCD potential in Sec.~\ref{Sec_II'}. 
Details of our lattice setup are given in Sec.~\ref{Sec_III}.
Numerical results and discussions on eigenfunctions, effective energies and the overlapping factors for $\Omega_{3Q}\Omega_{3Q}$ are presented in Sec.~\ref{Sec_IV}. 
Sec.~\ref{Sec_V} is devoted to summary and concluding remarks.
The analyses for the higher excited states are given in Appendix.~\ref{ApA}.

\section{The HAL QCD Method}\label{Sec_II}
The equal time NBS amplitude for two Omega baryons with energy $E$ is defined in the Euclidean spacetime as
\begin{equation}
\Psi_E(\bm r)e^{-Et} = \frac{1}{Z_\Omega}\sum_{\bm x}\braket{0|\hat\Omega(\bm x+\bm r, t)\hat\Omega(\bm x, t)|2\Omega,E},
\label{eq:NBS-wf}
\end{equation}
where $\hat\Omega$ is a local interpolating operator for $\Omega \equiv \Omega_{3Q}$  ($Q=s$ and $c$) whose explicit form can be found in Refs.~\cite{Gongyo2018, Lyu2021}.
$Z_\Omega$ is the wave function renormalization factor and $\ket{2\Omega,E}$ is the $2\Omega$ eigenstate with the center of mass energy $E=2\sqrt{m_{\Omega}^2 + k^2}$. 
  
Using Eq.(\ref{eq:NBS-wf})  and  the Haag-Nishijima-Zimmermann reduction formula for composite particles~\cite{Zimmermann1987}, the interaction between baryons can be identified as the energy-independent non-local potential $U(\bm r, \bm r')$ in the equal-time NBS equation ~\cite{Ishii2007, Aoki2010}:
 \begin{equation}
  \frac{1}{m_{\Omega}} (k^2+\nabla^2) \Psi_E(\bm r) = \int d\bm r' U(\bm r, \bm r')\Psi_E(\bm r').
 \label{eq:HAL-eq1}
\end{equation}

In the time-dependent HAL QCD method~\cite{Ishii2012, Aoki:2012bb}, the spacetime correlation function $R(\bm r, t)$ is introduced as a linear superposition of the NBS wave functions  below the inelastic threshold;
\begin{equation}
\begin{split}
R(\bm r, t) &= {\sum_{\bm x}\braket{0|\hat\Omega(\bm x+\bm r, t)\hat\Omega(\bm x, t)\overline{\mathcal J}(0) |0}}/{(Z_\Omega e^{-2m_\Omega t})}\\
&= \sum_n a_n\Psi_{E_n}(\bm r)e^{-(\Delta E_n) t}+O(e^{-(\Delta E^*)t}).
\end{split}
\end{equation}
Here $E_n$ is the eigenenergy of the $n$-th elastic scattering state in a finite box, and $ \Delta E_n = E_n - 2m_\Omega$ with $\Delta E_n  \le  \Delta E^*$: The inelastic threshold is denoted by  $\Delta E^* \sim \Lambda_\mathrm{QCD}\sim300~\mathrm{MeV}$.
The overlapping factors $a_n=\braket{2\Omega,E_n|\overline{\mathcal J}(0) |0}$ depend on the choice of the   source operator $\overline{\mathcal J}(0) $ at $t=0$.
If  $t\gg(\Delta E^*)^{-1}\sim0.7$ fm, effects from the inelastic states are exponentially suppressed as  $O(e^{-(\Delta E^*)t})$, so that one can rewrite Eq.(\ref{eq:HAL-eq1})  into the time-dependent HAL QCD equation,
\begin{equation}
 \left(\frac{1}{4m_{\Omega}}\frac{\partial^2}{\partial t^2}-\frac{\partial}{\partial t}-H_0\right)R(\bm r, t)=\int d\bm{r}'U(\bm r, \bm r')R(\bm r', t),
\label{eq:TDHAL}
\end{equation}
with $H_0 = -\nabla^2/m_{\Omega}$.  
The advantage of this equation is that we do not need to separate each eigenenergy $E_n$ to extract the potential $U$.
Since the potential is spatially localized in QCD by construction, its finite volume correction is exponentially suppressed for  large lattice volume. 
It has been also demonstrated  in Refs.~\cite{Iritani2019Jhep,Iritani2019PRD} that observables (such as the phase shifts) do not depend on the choice of source operators $\overline{\mathcal J}(0)$ (either wall-type source or smeared source) by taking  the $\Xi\Xi$ system as an example with the lattice volumes $\simeq (3.6, 4.3, 5.8 \ {\rm fm})^3$  and   $m_{\pi}\simeq$ 510 MeV, as long as the non-locality of the potential is well approximated by the derivative expansion. 
    
We note that it is practically useful to make a derivative expansion of the  non-local potential as 
\begin{equation}
U(\bm r, \bm r')=V({\bm r})\delta(\bm r-\bm r')+\sum_{n=1}V_{2n}(\bm r)\nabla^{2n}\delta(\bm r-\bm r').
\end{equation}
Then the leading-order (LO)   potential is obtained as 
\begin{equation}\label{Eq_LOV}
V({\bm r})=R^{-1}(\bm r,t) \left(\frac{1}{4m_{\Omega}}\frac{\partial^2}{\partial t^2}-\frac{\partial}{\partial t}-H_0\right)R(\bm r, t).
\end{equation}

\section{Optimized Sink Operators}\label{Sec_II'}
Once the HAL QCD potential  is obtained from the LQCD simulation of $R({\bm r},t)$ in  Eq.(\ref{eq:TDHAL}), 
one can calculate  eigenfunctions and eigenenergies in a finite lattice box by solving  Eq.(\ref{eq:HAL-eq1}).
This enables us to construct a set of optimized interpolating operators which couple strongly to each eigenstate, as originally proposed in Ref.~\cite{Iritani2019Jhep}.  
We now apply this idea to find optimized sink operators in the  present systems.

Let us first rewrite Eq.(\ref{eq:HAL-eq1}) in a three-dimensional  lattice box by introducing the Hamiltonian $H$ with the discretized Laplacian $H_0$ and the non-local potential $U(\bm r, \bm r')$; 
\begin{equation}\label{Eq_H}
H=H_0 + U, \quad H\Psi_{E_n}=\varepsilon_n\Psi_{E_n},
\end{equation}
where $\varepsilon_n$ is related to $\Delta E_n$ as  
\begin{equation}\label{Eq_E0}
\Delta E_n=2(\sqrt{\varepsilon_n m_\Omega+m_\Omega^2}-m_\Omega).
\end{equation}
From the $n$-th NBS wave function $\Psi_{E_n}(\bm r)$ for Hermitian matrix $U(\bm r, \bm r')$, 
\footnote{For non-Hermitian $U(\bm r, \bm r')$,  $\Psi_{E_n}^\dagger(\bm r)$ in Eqs.(\ref{eq:Sn},\ref{Eq_R_prj})   should be replaced by
$\sum_{m} N^{-1}_{nm}  \Psi_{E_{m}}^\dagger(\bm r)$ with $N_{nm}= \sum_{\bm r}  \Psi_{E_n}^\dagger(\bm r) \Psi_{E_{m}}(\bm r)$ being the norm kernel~\cite{Aoki2010,Aoki:2012bb}.}
 one may construct an optimized sink operator as a projection to each $n$-th state,
\begin{eqnarray}\label{eq:Sn}
 S_n(t) =  \sum_{\bm r}\Psi_{E_n}^\dagger(\bm r)\left[\sum_{\bm x}\hat\Omega(\bm x+\bm r, t)\hat\Omega(\bm x, t)\right].
\end{eqnarray}
This  is expected to couple primarily  to $\ket{2\Omega,E_n}$ for large $t$.
Two-baryon temporal correlation after the projection reads
\begin{eqnarray}\label{Eq_R_prj}
 R_n(t)&=&\sum_{\bm r}\Psi_{E_n}^\dagger(\bm r)R(\bm r, t),\nonumber \\
            &=&   a_n e^{-(\Delta E_n) t}+O(e^{-(\Delta E^*)t}),
\end{eqnarray}
which can be  used to extract effective two-baryon energy for each $n$ as 
\begin{equation}\label{Eq_E}
\Delta E^\mathrm{eff}_n(t)=\frac{1}{a}\ln\left[\frac{R_n(t)}{R_{n}(t+1)}\right].
\end{equation}
This  quantity should have plateau structure as a function of $t$ and approach to  $\Delta E_n$ in Eq.(\ref{Eq_E0}) for large $t$.
Note that the  projection to each eigenstate is possible only if we have spatial information of the correlation function $R({\bm r},t)$.
For similar attempts based on the spatial information of the correlation function in different physical contexts, see Refs.~\cite{Umeda2000,Larsen2019,Chen2021}.
  
In the case where only the LO potential $V(\bm r)$ in Eq.(\ref{Eq_LOV}) is available,  eigenfunction $\psi_n(\bm r)$ from the LO Hamiltonian  $H_{\rm LO}\equiv H_0 + V$ is an approximation of  the exact $n$-th NBS wave function $\Psi_{E_n}(\bm r)$. 
Hence resultant quantities in Eqs.(\ref{Eq_E0})-(\ref{Eq_E}) are approximate ones.
Then, whether  $\Delta E^\mathrm{eff}_n(t)$ has a plateau and approaches to Eq.(\ref{Eq_E0}) at large $t$ provide a confidence test on  the  truncation of the derivative expansion of $U$.

For later purpose, let us introduce the unprojected temporal correlation, $R(t)= \sum_{\bm r} R({\bm r}, t)$, which can be decomposed as 
\begin{eqnarray}\label{Eq_R_nprj}
 R(t)=\sum_n b_n e^{-(\Delta E_n) t}+O(e^{-(\Delta E^*)t}),
\end{eqnarray}
where $b_n = a_n \sum_{\bm r} \Psi_{E_n}({\bm r})$. 
Apparently, $R(t)$ is a superposition of different states and approaches to the ground state only when 
$t \gg ({\Delta E_1 - \Delta E_0})^{-1} \sim m_{\Omega} \left({La}/{2\pi}\right)^2$
which becomes unrealistically large for large volume and/or heavy hadrons. 
(Here $L$ is the number of lattice sites in one spatial direction  and $a$ is the lattice spacing.)
Note that $R(t)$ may create a ``fake plateau" for intermediate values of $t$ due to excited state contaminations~\cite{Iritani2016,Iritani:2017rlk,Iritani2019Jhep}.  

\section{Lattice setup}\label{Sec_III}

Numerical data used in this study are obtained from the ($2+1$)-flavor gauge configurations with Iwasaki gauge action at $\beta=1.82$ and nonperturbatively $O(a)$-improved Wilson quark action with stout smearing at nearly physical quark masses~\cite{Ishikawa2016}.
The relativistic heavy quark action~\cite{Aoki2003} for the charm quark~\cite{Namekawa2017} is used to remove cutoff errors associated with the charm quark mass up to next-to-leading order.
We have  $a\simeq0.0846$~fm ($a^{-1}\simeq2.333$~GeV) and $L=96$, which leads to $La\simeq8.1$~fm.
The pion, kaon,  $\Omega_{3s}$, and $\Omega_{3c}$ masses read ($m_\pi$, $m_K$, $m_{\Omega_{3s}}$, $m_{\Omega_{3c}}$) $\simeq$ (146, 525, 1712, 4796~MeV). 
The correlation functions are calculated by the unified contraction algorithm~\cite{Doi2013}.
For the source operator  $\overline{\mathcal J}(0) $, we use the wall-type  with the Coulomb gauge fixing.  

In order to increase statistics, forward and backward propagations are averaged, the hypercubic symmetry on the lattice (4 rotations) are utilized, and multiple measurements are performed by shifting the source position along the temporal direction. 
The total measurements for $2\Omega_{3s}$ ($2\Omega_{3c}$) amounts to 307,200 (896), where $2\Omega_{3Q}$ is a shorthand notation for $\Omega_{3Q}\Omega_{3Q}$. Note that current statistics for $2\Omega_{3s}$ are twice as those in Ref.~\cite{Gongyo2018}.
The statistical errors  are evaluated by the jackknife method throughout this paper.
For more numerical details, see Refs.~\cite{Gongyo2018, Lyu2021}.


\section{Numerical Results}\label{Sec_IV}

\begin{figure}[thb]
    \centering
    \includegraphics[width=8cm]{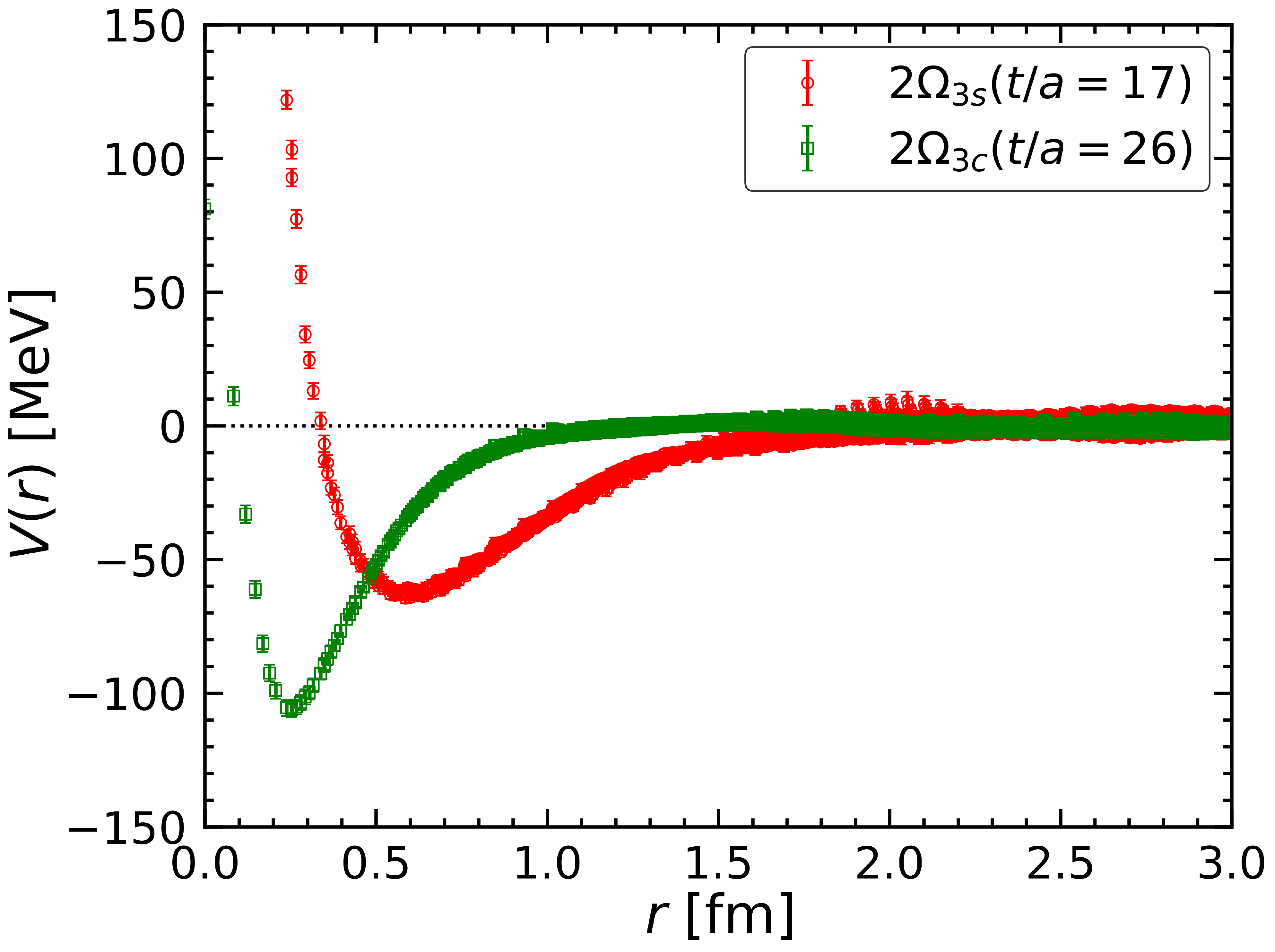}
    \caption{(Color online). The one-dimensional projection of the LO potential in the $A_1$-rep, $V(r) \equiv \left. V({\bm r})\right|_{r = |{\bm r}|}$,  for $2\Omega_{3s}$~\cite{Gongyo2018} at Euclidean time $t/a= 17$ (red circles), and for $2\Omega_{3c}$~\cite{Lyu2021} at Euclidean time $t/a= 26$ (green squares).
    }\label{Fig1}
\end{figure}

\begin{figure*}[thb]
    \centering
    \includegraphics[width=14cm]{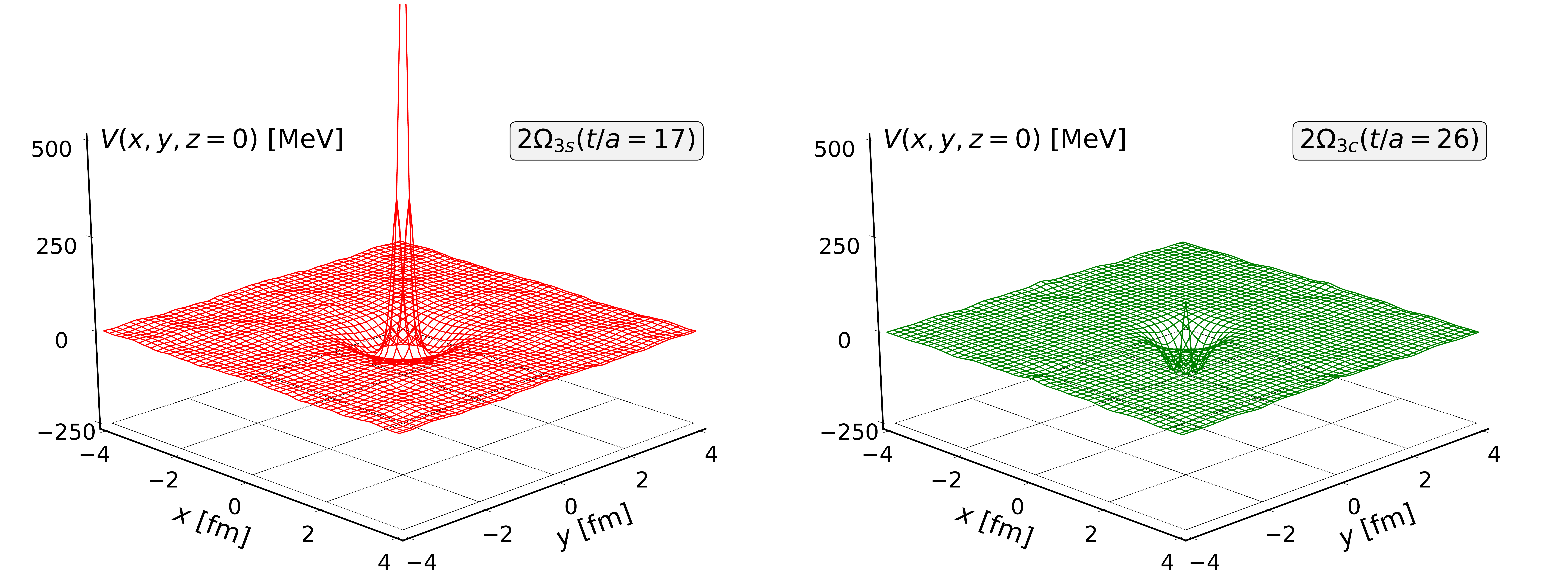}
    \caption{(Color online). The LO potential $V(x,y,z=0)$ in the $A_1$-rep for $2\Omega_{3s}$~\cite{Gongyo2018} at Euclidean time $t/a= 17$ (left), and for $2\Omega_{3c}$~\cite{Lyu2021} at Euclidean time $t/a= 26$ (right).
    }\label{Fig2}
\end{figure*}

In Fig.~\ref{Fig1}, we show the one-dimensional projection of the LO potential on the three-dimensional lattice box, $V(r) \equiv \left. V({\bm r})\right|_{r = |{\bm r}|}$, obtained by the time-dependent HAL QCD method in the $A_1$ representation of the cubic group $SO(3,\mathbb{Z})$ ($A_1$-rep in short) for $2\Omega_{3s}$~\cite{Gongyo2018} at $t/a=17$, and for $2\Omega_{3c}$~\cite{Lyu2021} at $t/a=26$.
These Euclidean times are chosen such that they are large enough to suppress contaminations from excited states in the single-baryon correlator and simultaneously small enough to avoid exponentially increasing statistical errors.  
The potential for  $2\Omega_{3c}$ is shorter ranged with smaller repulsive core than that for  $2\Omega_{3s}$.
In both systems, the LO potentials $V({\bm r})$ are localized around the origin and are approximately spherical functions as shown in  Fig.~\ref{Fig2} for $V(x,y,z=0)$.
Also, $V(r)$ in Fig.~\ref{Fig1} can be well-fitted by the three-range Gaussians, $V_{\rm fit}(r) = \sum_{i=1,2,3} \alpha_i \exp(-\beta_i r^2)$~\cite{Gongyo2018,Lyu2021}.   
Due to the large cancellation between the medium-range attraction and the short-range repulsion, only one loosely bound state appears  in the infinite volume ($L=\infty$) for each of $2\Omega_{3s}$~\cite{Gongyo2018} and $2\Omega_{3c}$~\cite{Lyu2021}. 

\subsection{Eigenfunctions of $H_{\rm LO}$ on the lattice}\label{Sec_IV-1}
\begin{figure}[t]
\centering
\includegraphics[width=8cm]{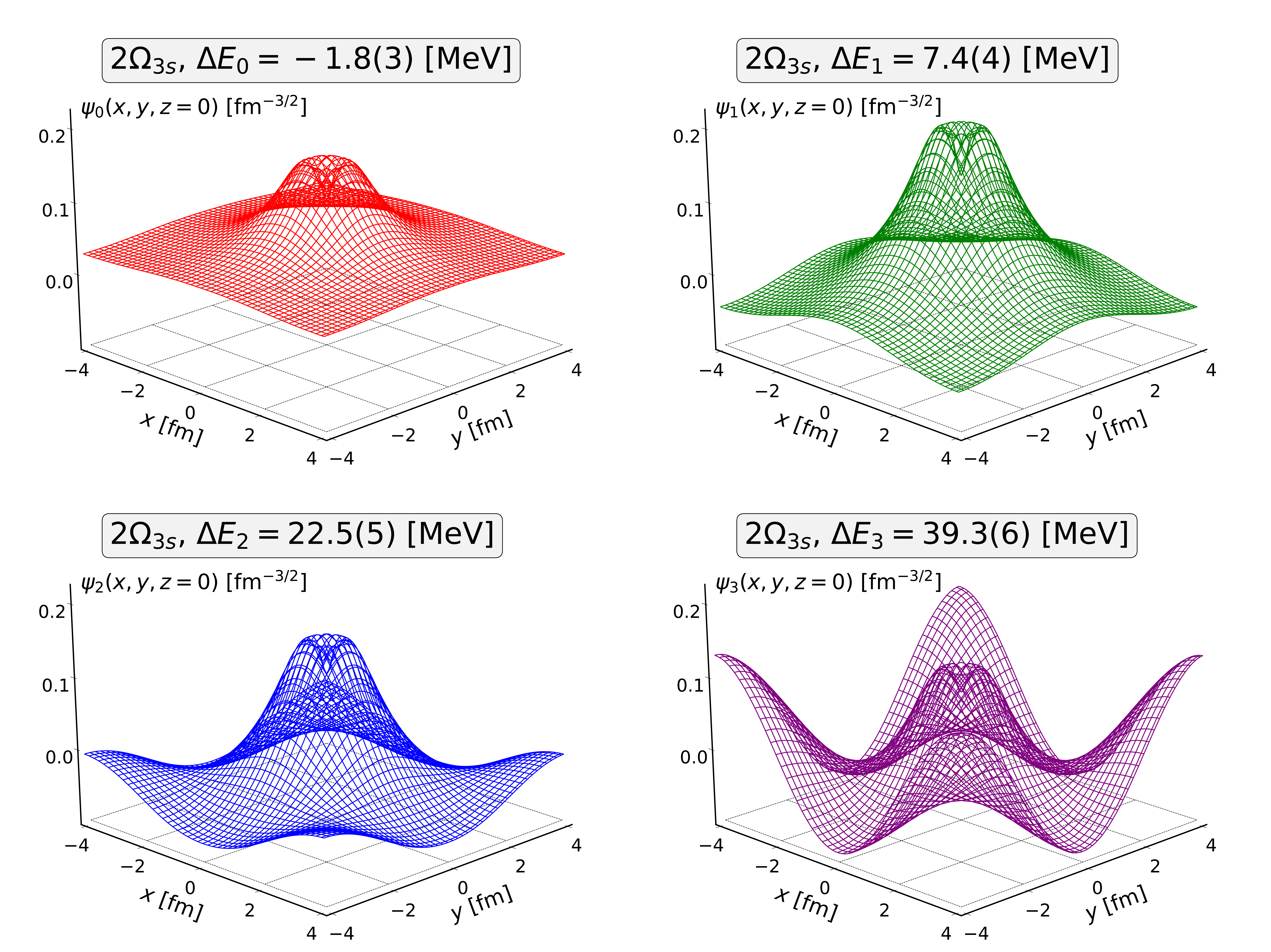}
\includegraphics[width=8cm]{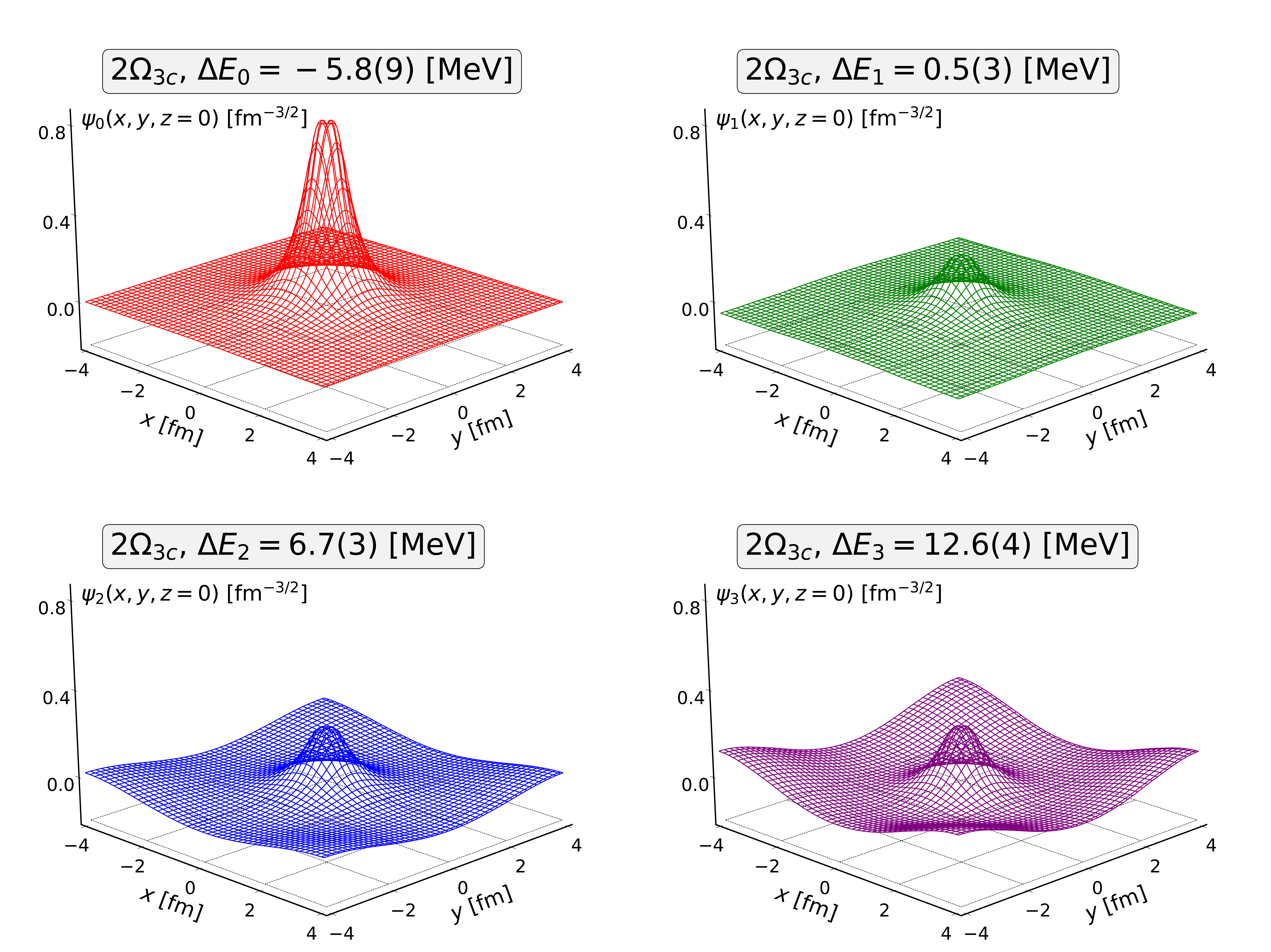}
\caption{(Color online). The first four eigenfunctions $\psi_n(x,y,z=0)$ of the LO Hamiltonian $H_{\rm LO}$ in the $A_1$-rep for $2\Omega_{3s}$ (upper four panels) and for $2\Omega_{3c}$ (lower four panels). They are normalized as $\sum_{\bm r}|\psi_n(\bm r)|^2=1$ with a convention $\psi_n(\bm 0)>0$. The red, green, blue, and purple wireframes correspond to $\psi_n(x,y,z=0)$ with $n=0$, $1$, $2$, and $3$, respectively. The corresponding eigenenergy $\Delta E_n$ is shown at the top of each panel. }
\label{Fig3}
\end{figure}

Using the LO potential $V({\bm r})$, the discretized Hamiltonian $H_{\rm LO} \equiv H_0+V$ can be diagonalized on the three-dimensional lattice box with the size $(La)^3 \simeq (8.1 {\rm fm})^3$ and $L=96$.
Under the periodic boundary condition, we thus obtain the eigenenergies $\Delta E_n$ and associated eigenfunctions $\psi_n(\bm r)$ in the $A_1$-rep. 
Note that $\psi_n$ here represents an approximation of the exact $n$-th NBS wave function $\Psi_{E_n}$ associated with the LO potential $V$ as discussed in Sec.~\ref{Sec_II'}.
The eigenfunctions are normalized $\sum_{\bm r}|\psi_n(\bm r)|^2=1$ with a convention $\psi_n(\bm 0)>0$.
Shown in Fig.~\ref{Fig3} are the first four states $(n=0,1,2,3)$,  $\psi_{n}(x,y,z=0)$, together with $\Delta E_{n}$. 
The eigenfunctions are distorted by the boundary condition and have only discrete rotational symmetry.
Therefore those in the $A_1$-rep not only contain component with angular momentum $l=0$ but also components with $l=4, 6, \cdots$.
Such a mixing becomes prominent as $r$ and/or $n$ increase.

\begin{figure}[t]
\centering
\includegraphics[width=8cm]{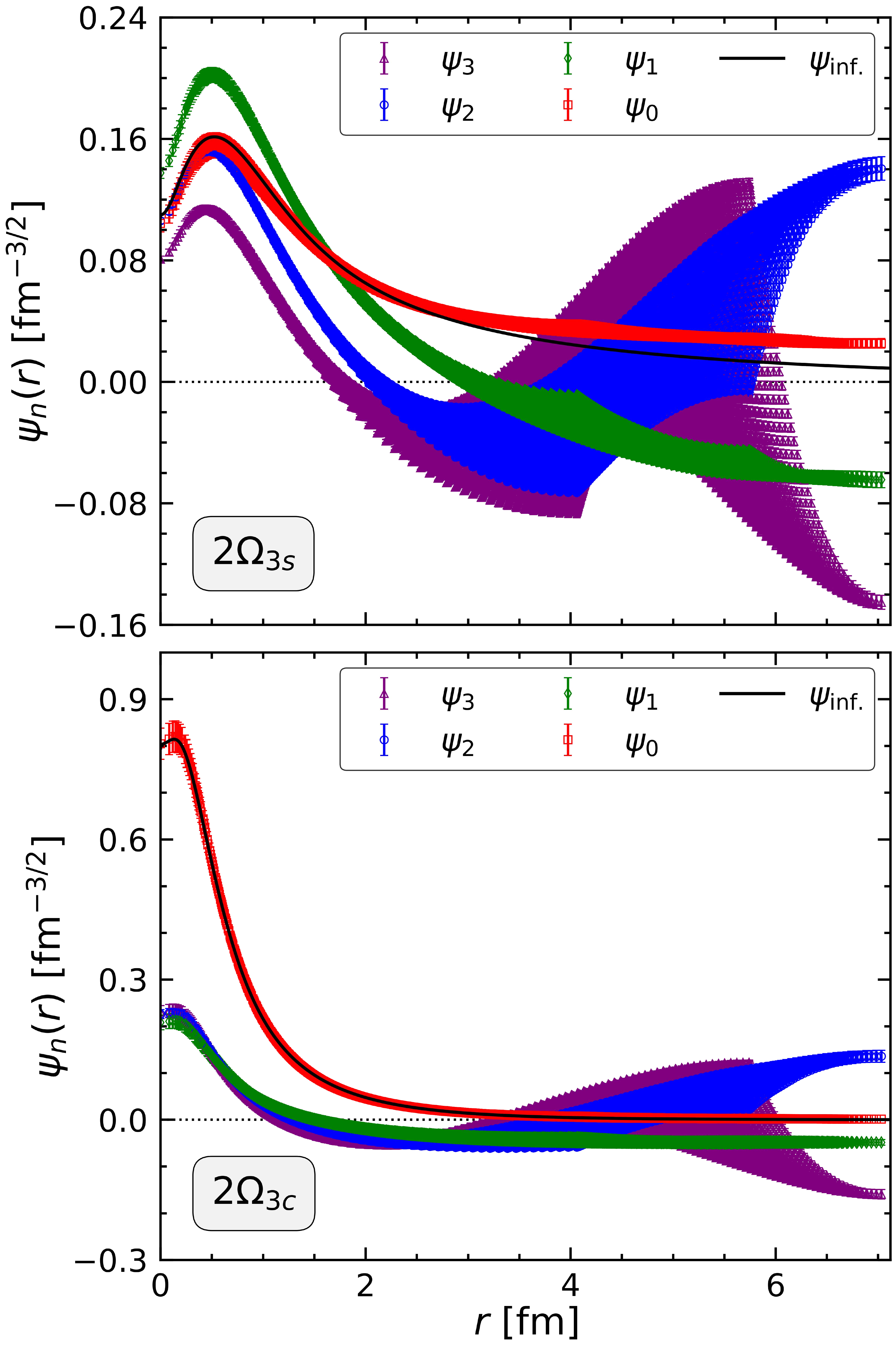}
\caption{(Color online). The one-dimensional projection of eigenfunctions of $H_{\rm LO}$ in the $A_1$-rep on the three-dimensional lattice box, $\psi_n(r)\equiv \left. \psi_n(\bm r)\right|_{r=|\bm r|}$ for $2\Omega_{3s}$ (upper panel) and for $2\Omega_{3c}$ (lower panel). The red squares, green diamonds, blue circles, and purple triangles correspond to $\psi_n(r)$ with $n=0$, $1$, $2$, and $3$ , respectively.
The black solid lines are the bound-state wavefunctions $\psi_\mathrm{inf.}(r)$ in the infinite and continuum space obtained by solving the Schr\"{o}dinger equation with $V_{\rm fit}(r)$.}
\label{Fig4}
\end{figure}

Shown in Fig.~\ref{Fig4} are the one-dimensional projection of the above eigenfunctions, $\psi_{n}(r) \equiv \left. \psi_{n}({\bm r})\right|_{r=|{\bm r}|}$ on the three-dimensional lattice box. 
Note that  the number of nodes of $\psi_n(r)$ is equal to $n$ as expected from the quantization condition given by the periodic boundary condition.   
The characteristic size of the ground state is smaller for the $2\Omega_{3c}$ than that for $2\Omega_{3s}$.
Shown together by the black solid lines are the bound-state eigenfunctions $\psi_{\rm inf.}(r)$ in the infinite ($L=\infty$) and continuous ($a=0$) space obtained by solving the Schr\"{o}dinger equation with $H^{\infty}_{\rm LO}=-\nabla^2/m_{\Omega}+V_{\rm fit}(r)$.
We find that $\psi_{\rm inf.}(r)$ and $\psi_0(r)$ are indistinguishable for $r< 3$ fm in both cases.

\subsection{Effective energies on the lattice}\label{Sec_IV-2}

\begin{figure}[t]
\centering
\includegraphics[width=8cm]{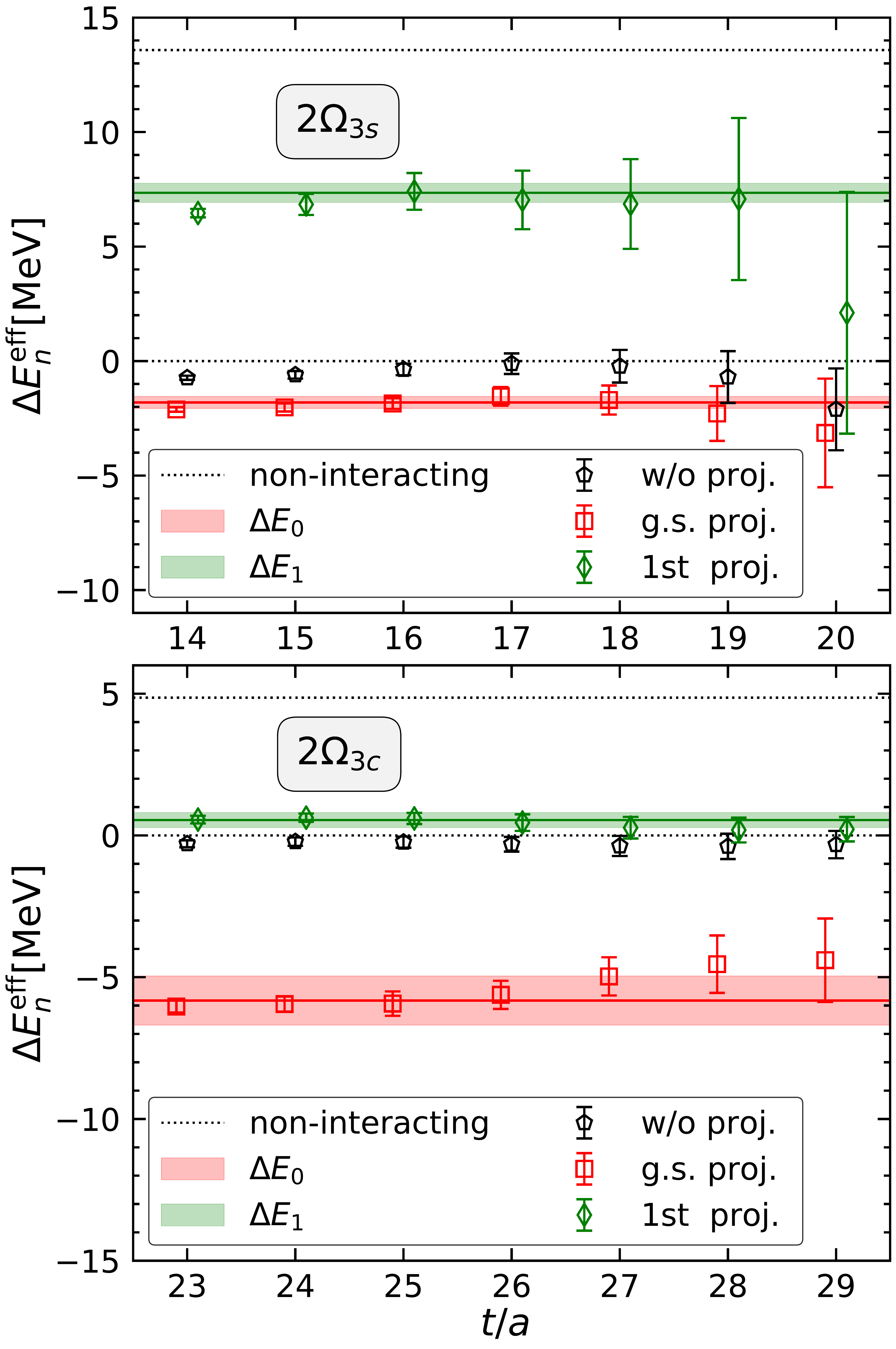}
\caption{(Color online). The effective energies $\Delta E^\mathrm{eff}_n(t)$ in Eq.~(\ref{Eq_E}) obtained from the projected temporal correlators $R_n(t)$ with $n=0$ (red squares) and $1$ (green diamonds) for $2\Omega_{3s}$ (upper panel) and for $2\Omega_{3c}$ (lower panel). The red (green) bands show $\Delta E_0$ ($\Delta E_1$) calculated from the LO Hamiltonian $H_{\rm LO}$, while the black dotted lines are $\Delta E_{0,1}$ for a non-interacting system. The black pentagons represent the effective energies extracted from the temporal correlations without projection, $R(t)=\sum_{\bm r}R(\bm r,t)$. 
}
\label{Fig5}
\end{figure}

Let us now utilize the eigenfunctions $\psi_n({\bm r})$ to define optimized two-baryon sink operators $S_n(t)$ in Eq.(\ref{eq:Sn}) 
 and evaluate temporal correlators $R_n(t)$ in Eq.(\ref{Eq_R_prj}) to derive the effective energy $\Delta E^\mathrm{eff}_n(t)$.
Shown in Fig.~\ref{Fig5} by the open squares and  open diamonds as a function of $t/a$ 
are $\Delta E^\mathrm{eff}_n(t)$ for $n=0$ and  $1$, respectively.  
To be consistent with the Euclidean time employed to extract $V({\bm r})$, 
 we choose $t/a = 17\pm3$  ($26\pm 3$) for $2\Omega_{3s}$ ($2\Omega_{3c}$).
The colored bands in Fig.~\ref{Fig5} are $\Delta E_n$ obtained from $H_{\rm LO}$ as discussed in Sec.\ref{Sec_IV-1}.
Open pentagons are the ``naive''  ground-state energy,
$\widetilde{\Delta E}(t)  =\frac{1}{a}\ln\frac{R(t)}{R(t+1)},$ obtained from the unprojected temporal correlations, $R(t)=\sum_{\bm r}R(\bm r, t)$.  
  
By comparing $\Delta E^\mathrm{eff}_n(t)$, $\Delta E_n$  and $\widetilde{\Delta E}(t) $ in Fig.~\ref{Fig5}, we find the following points:
\begin{itemize}
\item[(i)] Each effective energy ($\Delta E^\mathrm{eff}_{0}(t)$ and  $\Delta E^\mathrm{eff}_{1}(t)$) has its own plateau for both $2\Omega_{3s}$ and $2\Omega_{3c}$. 
In addition, good agreements between $\Delta E^\mathrm{eff}_{0,1}(t)$ and $\Delta E_{0,1}(t)$ are found.
\item[(ii)]  The above observations indicate that different eigenstates are properly separated by the projection, and the uncertainty from the derivative expansion of $U$ is within the statistical errors.
Otherwise, eigenfunctions $\psi_{0,1}({\bm r )}$ obtained from $V$ would be different from the exact NBS wave functions $\Psi_{E_{0,1}}({\bm r})$.
 
\item[(iii)] Although $ \widetilde{\Delta E}(t) $ looks stable, its magnitude  does not agree with the
 ground state energy $\Delta E_0$. 
This is the typical  ``fake plateaux''  behavior due to the contamination from the excited states in $R(t)$.
This phenomenon for two-baryon systems was extensively studies in Refs.~\cite{Iritani2016,Iritani:2017rlk,Iritani2019Jhep}.
\end{itemize}

The same analyses for $n=2$ and $3$ are given in Appendix~\ref{ApA} where the agreement 
between $\Delta E^\mathrm{eff}_{n}(t)$ and $\Delta E_{n}$ is also observed within the statistical errors.
  
\subsection{Overlapping factors}\label{Sec_IV-3}

\begin{figure}[t]
\centering
\includegraphics[width=8cm]{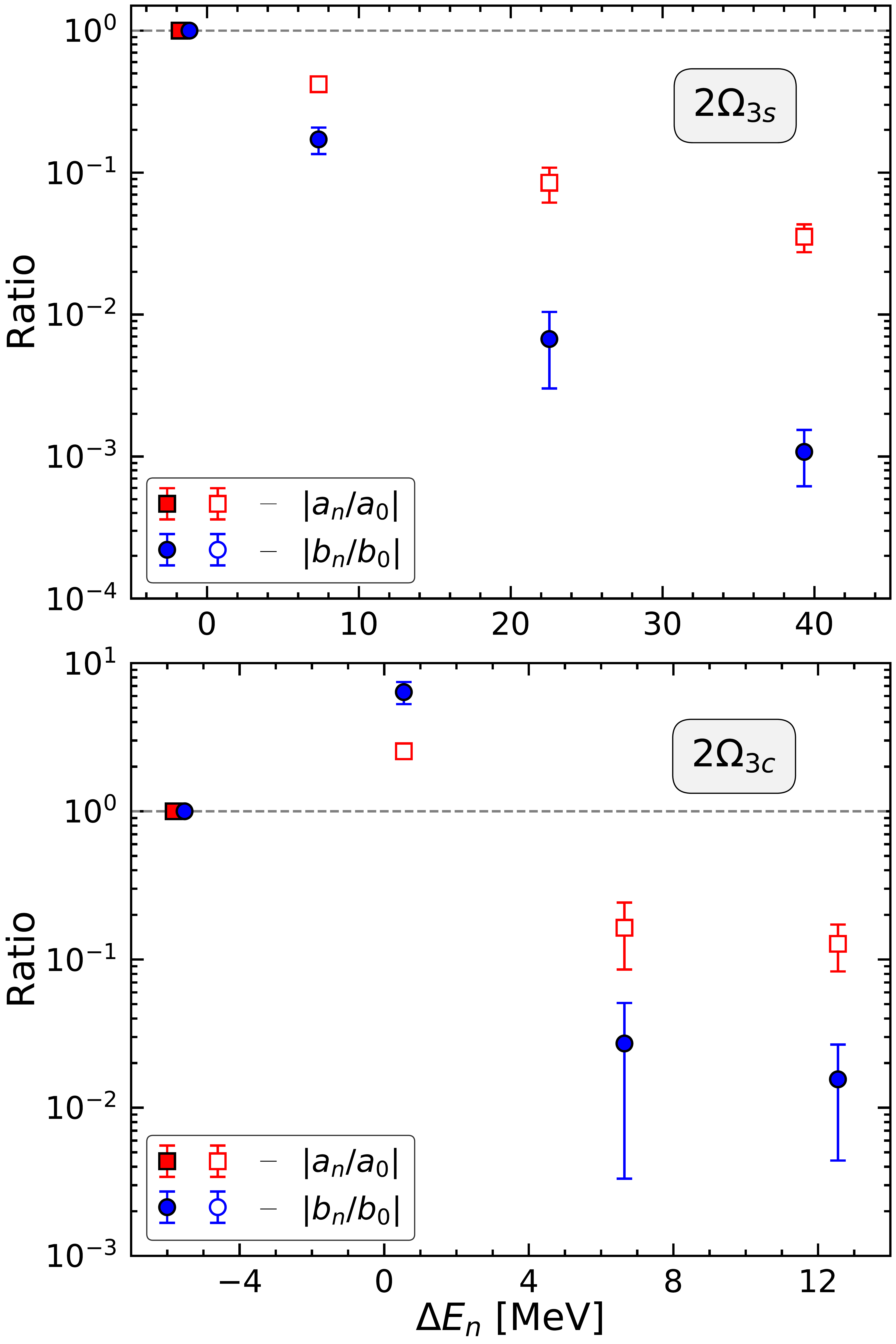}
\caption{(Color online). The ratio $|a_n/a_0|$ (red squares) and $|b_n/b_0|$ (blue circles) with $n=0$, 1, 2, and 3 as functions of $\Delta E_n$ for both $2\Omega_{3s}$ (upper panel) and $2\Omega_{3c}$ (lower panel).
The solid (open) symbols show the positive (negative) sign for the ratio.}
\label{Fig6}
\end{figure}

The coefficients $a_n$ in  Eq.(\ref{Eq_R_prj}) and $b_n$ in Eq.(\ref{Eq_R_nprj}) represent the overlapping strengths of the projected and unprojected sink operators to the state created by wall-type source, respectively. 
By using the eigenfunctions obtained in Sec.~\ref{Sec_IV-1},  these coefficients can be calculated as $a_n= \sum_{{\bm r}} \psi^{\dagger}_n({\bm r}) R({\bm r},t) e^{(\Delta E_n)  t}$ and $b_n= a_n \sum_{{\bm r}} \psi_n({\bm r})$~\cite{Iritani2019Jhep}.   
Note that the source and sink operators are not Hermitian conjugate with each other, so that the overlapping coefficients are not necessarily positive. 
To see the relative magnitude of the coefficients, we plot $|a_n/a_0|$ and $|b_n/b_0|$ versus $\Delta E_n$  in  Fig.~\ref{Fig6}.  
These ratios for $2\Omega_{3s}$ ($2\Omega_{3c}$) are evaluated at the center of the plateau, $t/a=17$ ($t/a=26$),  in  Fig.~\ref{Fig5}. 

From the red squares for $2\Omega_{3s}$ in Fig.~\ref{Fig6},  one finds that the wall-type source couples primarily to the $n=0$ state, secondly to the $n=1$ state with $|a_1/a_0|\simeq 0.4$, and negligibly to the higher states  $|a_{2,3}/a_0| < 0.1$.
On the other hand, the red squares for $2\Omega_{3c}$ indicate that the wall-type source couples primarily to the $n=1$ state with $|a_1/a_0|\simeq 3$. 
This is due to the fact that the size of the $n=0$ state for $2\Omega_{3c}$ is  rather small, so that its coupling to the extended wall-type source  is weak.
As we mentioned in Sec.~\ref{Sec_II}, the advantage of the HAL QCD method is that the decomposition of $R({\bm r},t)$ into  each eigenstate is not required to derive the potential, since the potential is independent of $n$ as long as the system is in the elastic region.
Present results show that the interaction potentials can be extracted reliably  in the time-dependent HAL QCD method with the derivative expansion for both $2\Omega_{3s}$ and $2\Omega_{3c}$, regardless whether the ground state or the (first) excited state dominates the two-baryon correlation function.
  
Let us now turn to the discussion on the unprojected temporal correlation $R(t)$ which produces  the fake plateau, i.e.  the black pentagons in  Fig.~\ref{Fig5}.
From $|b_1/b_0|$ in Fig.~\ref{Fig6}, one may evaluate the value of $t/a$ above which the true ground state saturation is reached by $\widetilde{\Delta E}(t)$.
Let us demand the systematic error in the effective energy due to the first excited state contamination is bounded,
 $\left|\frac{\widetilde{\Delta E}(t)-\Delta E_0}{\Delta E_0}\right|< \varepsilon $. 
Utilizing $\widetilde{\Delta E}(t)=\ln \frac{R(t)}{R(t+1)}$ with $R(t)=\sum_{i=0,1}b_ie^{-(\Delta E_i)t}$, the above condition gives $t>\frac{1}{\Delta E_1 - \Delta E_0} \ln\left[\frac{b_1}{b_0}\frac{ 1-e^{ (1-\varepsilon) (\Delta E_0)-\Delta E_1}}{-1+e^{-\varepsilon(\Delta E_0)}}\right]$ in the case of $b_1/b_0>0$ and $\Delta E_{0}<0$. 
Taking $\varepsilon =0.1$ and using $\Delta E_{0,1}$ in Fig.~\ref{Fig3}, and $b_1/b_0 \simeq 0.2$ ($6$) for $2\Omega_{3s}$ ($2\Omega_{3c}$), we need to have $t/a > 580$ for $2\Omega_{3s}$ and $t/a > 1500$ for $2\Omega_{3c}$, respectively.
However, $R(t)$  for such large $t$ would suffer from exponentially large  statistical errors~\cite{Lepage1989}.

\section{Summary and concluding remarks}\label{Sec_V}

In this paper, we explored the idea of the optimized interpolating operators 
originally proposed in~\cite{Iritani2019Jhep} on the basis of the time dependent HAL QCD method.   
To reduce the statistical errors, we considered heavy dibaryons $\Omega_{3Q}\Omega_{3Q}$ ($Q=s, c$) in the $^1S_0$ channel and extracted the leading-order HAL QCD potential $V({\bm r})$ localized in space.
It was then used to obtain the eigenfunctions and  eigenenergies on a finite lattice box with the periodic boundary condition.
The eigenfunctions $\psi_n(\bm r) $ were then used to construct the projected two-baryon sink operators $S_n(t)$ such that they couple predominantly to each $n$-th state.
The temporal correlations  $R_n(t)$ with such optimized sink operators are used to extract corresponding effective energies $\Delta E^\text{eff}_n(t)$ which were found to have plateau structure for each $n$.
Moreover, they agree quantitatively with $\Delta E_n$ calculated by solving the Schr\"{o}dinger equation on the lattice with $V({\bm r})$.
Such a feature provides an indirect evidence that the leading-order potential $V$ is a good approximation of the 
non-local potential $U$ for $2\Omega_{3Q}$ systems within the statistical errors.
It also implies that  the baryon-baryon interaction potential can be reliably extracted from the time-dependent HAL QCD method, no matter whether the two-baryon correlation function is dominated by the ground state or the (first) excited state, while the analysis of the unprojected temporal correlation $R(t)$ leads to a wrong effective energy associated with a fake plateau.

Although we applied our projection only to the sink operators in this paper, one may apply the idea to the source operators as well to further improve the stability and accuracy of $\Delta E^{\rm eff}_n(t)$.  
Such an approach can also provide an optimized operator basis ($S_n$) for the conventional variational method~\cite{Luscher:1990ck} to study hadron-hadron interactions through the matrix correlation,
$C_{nn'} (t) = \langle 0| S_n(t) S_{n'}^{\dagger} (0) |0 \rangle$.  Further Investigation along this line will be reported elsehwhere.
  

\begin{acknowledgments}
We thank the members of the HAL QCD Collaboration for stimulating discussions.
Y.L. thanks Xu Feng for helpful discussions. 
The lattice QCD data  used in this work was generated on  K and HOKUSAI at RIKEN, and HA-PACS at Univ.of Tsukuba.
We thank ILDG/JLDG \cite{ldg}, which serves as an essential infrastructure in this study.
This work was partially supported by HPCI System Research Project (hp120281, hp130023, hp140209, hp150223, hp150262, hp160211, hp170230, hp170170, hp180117, hp190103, hp200130 and hp210165),  the National Key R\&D Program of China (Contract Nos. 2017YFE0116700 and 2018YFA0404400), the National Natural Science Foundation of China (Grant Nos. 11935003, 11975031, 11875075, and 12070131001), the JSPS (Grant Nos. JP18H05236, JP16H03978, JP19K03879, and JP18H05407), the MOST-RIKEN Joint Project ``Ab initio investigation in nuclear physics'', ``Priority Issue on Post-K computer'' (Elucidation of the Fundamental Laws and Evolution of the Universe), ``Program for Promoting Researches on the Supercomputer Fugaku'' (Simulation for basic science: from fundamental laws of particles to creation of nuclei), and Joint Institute for Computational Fundamental Science (JICFuS).  
\end{acknowledgments}

\newpage

\appendix

\section{$n=2$ and $3$ cases}\label{ApA}
Here we show the same analyses along the line with the main text for higher excited states.

\begin{figure}[htb]
    \centering
    \includegraphics[width=8cm]{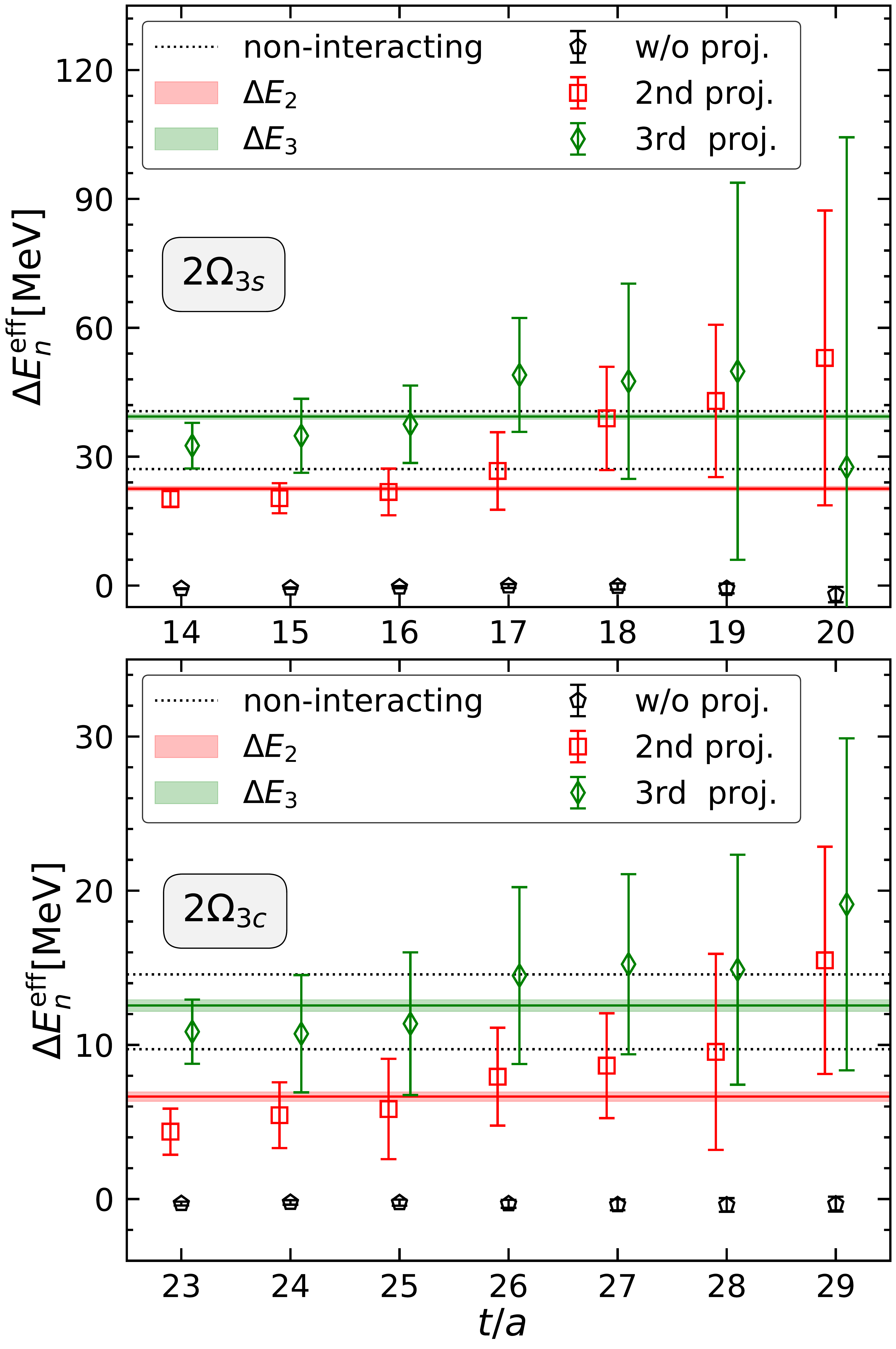}
    \caption{(Color online). The same as Fig.~\ref{Fig5}, but for $n=2$ and $3$.}
    \label{FigA}
\end{figure}

Fig.~\ref{FigA} shows the effective energies $\Delta E^\mathrm{eff}_n(t)$ obtained from the projected temporal correlators $R_n(t)$ with $n=2$ (red square) and $n=3$ (green diamond) for both $2\Omega_{3s}$ and $2\Omega_{3c}$. The colored bands are corresponding $\Delta E_n$ calculated from the LO Hamiltonian $H_{\rm LO}$. The black pentagons are effective energies obtained from the unprojected temporal correlators.
$\Delta E^\mathrm{eff}_{2,3}(t)$ are found to be consistent with $\Delta E_{2,3}$.
Large statistical errors in $\Delta E^\mathrm{eff}_{2,3}(t)$ are due to the fact that the contributions to $R(\bm r, t)$ from the second and third excited states are very small, as we can see in Fig.~\ref{Fig6}.

%


\end{document}